\documentclass{moriond}

\usepackage[numbers]{natbib}
\usepackage{wrapfig}
\usepackage{amssymb}
\usepackage{caption}
\usepackage{subcaption}

\def\be{\begin{equation}}
\def\ee{\end{equation}}
\def\bea{\begin{eqnarray}}
\def\eea{\end{eqnarray}}

\begin{document}

\vspace*{4cm}
\title{Search for Orphan Gamma-Ray Burst Afterglows with the Vera C. Rubin Observatory and the alert broker Fink}

\author{ M. Masson, J. Bregeon }

\address{Univ. Grenoble Alpes, CNRS, LPSC-IN2P3, 38000 Grenoble, France}

\maketitle

\abstracts{ Orphan gamma-ray burst afterglows are good candidates to learn more about the GRB physics and progenitors or for the development of multi-messenger analysis with gravitational waves. Our objective is to identify orphan afterglows in Rubin LSST data, by using the characteristic features of their light curves. In this work, we generated a population of short GRBs based on the Swift SBAT4 catalogue, and we simulated their off-axis afterglow light curves with \texttt{afterglowpy}. We then used the \texttt{rubin\_sim} package to simulate observations of these orphan afterglows with Rubin LSST and proceeded with the characterisation of orphan light curves by extracting a number of parameters. The same parameters are computed for the ELAsTiCC (Extended LSST Astronomical Time-series Classification Challenge) data set, a simulated alert stream of the Rubin LSST data. We then started to develop a machine learning filter able to discriminate orphan-like events among all the variable objects. We present here the performance of our filter as implemented in the Fink broker and tested on the ELAsTiCC data set and our own Rubin pseudo-observation simulations. }

\section{Introduction}

\subsection{Orphan gamma-ray burst afterglows}

Gamma-Ray Bursts (GRBs) are among the most energetic phenomena in the Universe. They are produced by relativistic jet from mergers of compact objects or the collapse of massive stars, and appear as short and intense flashes of gamma rays called prompt emission. The interaction of their blast wave with the medium surrounding the progenitor produces an afterglow that can be observed from a larger angle, in a wide range of the electromagnetic spectrum and during a longer time than the prompt emission. Viewed off-axis, this emission has a negligible gamma-ray flux and is hence called ``orphan GRB afterglow". Their properties make them good candidates to learn more about the GRB physics and progenitors or for the development of multi-messenger analysis, like in the case of GW 170817A \citep{Abbott_2017}, but today no orphan afterglow candidate has been confirmed yet. We will focus here on short GRBs with the long term goal of enabling more associations of gravitational wave signal with their electromagnetic counterpart, or, finding a new population of gravitational-wave standard sirens.

\subsection{The Vera C. Rubin Observatory and the alert broker Fink}

According to most theoretical models, optical orphan afterglows should be found as slow and faint transients. This is why the Vera C. Rubin Observatory, currently under construction in Chile, shall significantly improve their detection: thanks to its limiting nightly magnitude of 24.5 and its large field of view (9.6 deg²), it should be able to detect up to 50 orphans per year \citep{Ghirlanda_2015}. The Rubin Observatory will perform the Legacy Survey of Space and Time (LSST) over 10 years, a period during which about ten million alerts will be generated every single night. To deal with this huge alert stream, dedicated software called ``alert brokers" have been actively developped by different teams in the past years to categorise the alerts. We work on one of the official alert broker of Rubin LSST called Fink \citep{Moller_2020} in which we want to implement a filter to identify orphan afterglows in Rubin LSST data.

\subsection{A specific light curve}

With Rubin LSST, the identification of orphan afterglows candidates will only be based on their specific light curve. Considering a standard electron-synchrotron model as described in \citep{Sari_1998}, their light curve come as multiple power-law segments in $F_\nu \propto t^{-\alpha}$ and typically displays a fast rise and a slower decay, as illustrated in Figure \ref{lc}. Flux computations were conducted using the \texttt{afterglowpy} package developed by \citep{Ryan_2020}, with fluxes expressed in AB magnitudes.

To be observable, an orphan afterglow must maintain a magnitude lower than the Rubin Observatory limiting magnitude (24.5 in the r-band). Its observability is influenced in different ways by the model parameters, with our focus directed towards the most impacting ones: the isotropic equivalent energy of the jet $E_{iso}$, the ambient medium density $n_0$, the redshift $z$, the observer angle $\theta_{obs}$. Additionaly, we consider a structured jet with an energetic core width $\theta_c$. The determination of distributions for each parameter is needed to generate realistic estimations regarding the quantity and quality of detectable orphan afterglows.

\section{Simulation of a population of short Gamma-Ray Bursts}

\subsection{Population of short GRBs}

Distributions for the different parameters is not very well known, so to have a realistic population of short GRBs, we rely on prompt emission observations. We used the SBAT4 catalogue, a selected sample of short GRBs observed by the Swift satellite up to June 2013, detected within the 15-150 keV energy band. To ensure the catalogue completeness, only GRBs with a peak flux PF$_{64}$ $>$ 3.5 ph/s/cm² (PF$_{64}$ being the peak photon flux PF computed using the 15–150 keV Swift-BAT light curves binned with $\delta$t=64 ms) were included \citep{D_Avanzo_2014}.

The methodology employed here is outlined in \citep{Salafia_2023}. To create a SBAT4-like GRB population, we generate configurations of parameters with uniform distributions, calculate the flux that would be received by the Swift satellite in the BAT energy bands, and retain only those with a peak flux greater than 3.5 ph/s/cm². Since we are interested in the joint detection of orphan afterglow and gravitational wave signal, we focus here on the GRBs that are close enough to be detected by the LIGO-Virgo-Kagra detectors ($z < 0.1$ for binary neutron stars during the O5 run). We generate a population of $10^5$ short GRBs with $z<0.1$. 

To obtain the corresponding orphan afterglows, we compute the afterglow flux in the energy band of Fermi GBM (50-300 keV) and keep the afterglows that would be too faint to be detected by applying a cut at 0.5 ph/s/cm$^2$ (Fermi GBM detection limit). We add the criterion $\theta_c < \theta_{obs}$ to select only GRBs observed off-axis.

\subsection{Simulation of an observation}

For each orphan afterglow of the population, we computed the theoretical light curve in the r-band and retained only the ones that have a magnitude lower than 24.5. We then simulated their observation by using the \texttt{rubin\_sim} package, a tool provided by the Rubin LSST collaboration that contains a realisation of the scheduler simulation for the 10 years of LSST. 

\begin{figure}
    \centering
    \setlength{\belowcaptionskip}{-10pt}
    \begin{subfigure}[t]{0.48\textwidth}
        \centering
        \includegraphics[width=\linewidth]{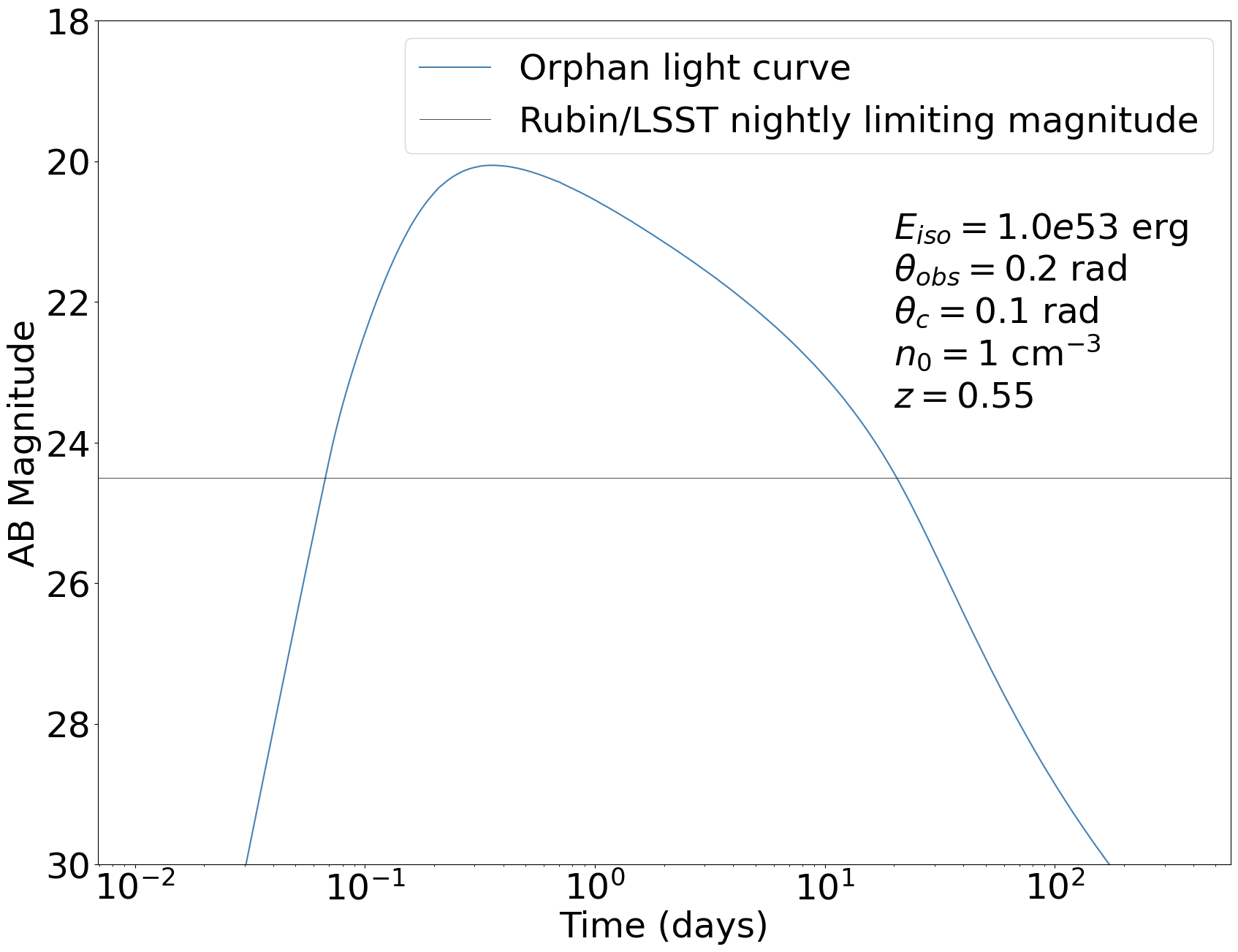}
    \end{subfigure}
    \begin{subfigure}[t]{0.48\textwidth}
        \centering
        \includegraphics[width=\linewidth]{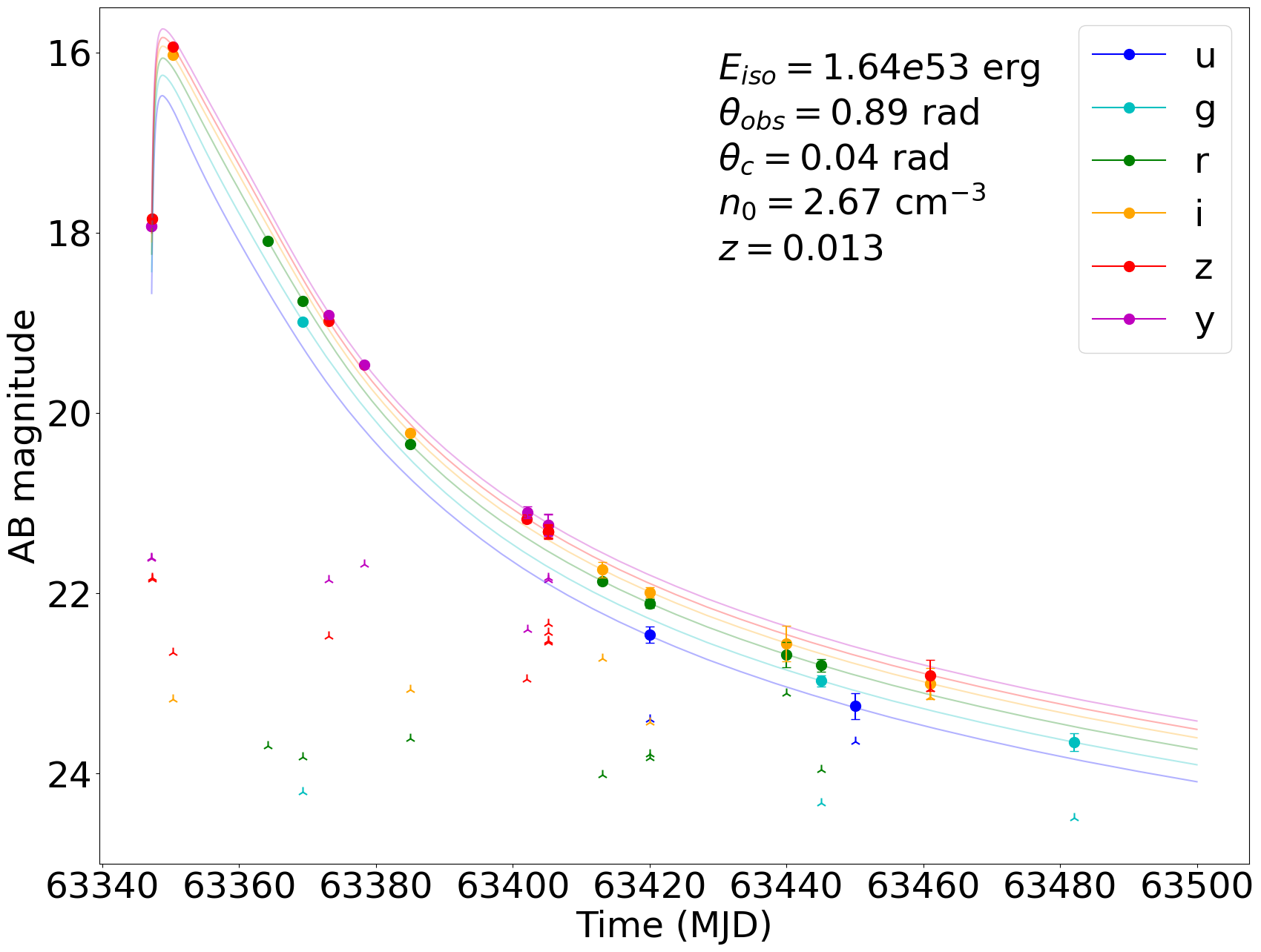}
    \end{subfigure}
    \caption{Orphan afterglow light curves. Left: theoretical light curve, computed with the \texttt{afterglowpy} package. Right: simulated observation of an orphan afterglow. Each ($\bullet$) marker is a detection point while the ($\curlywedge$) markers corresponds to the limiting magnitude at the time of the detection.}
    \label{lc}
\end{figure}

To simulate these observations, we randomly generated a date and a position in the sky, and assigned a configuration of parameters ($E_{iso}, \theta_{obs}, \theta_c, n_0, z$). That enabled us to know if the afterglow would fall within the field-of-view of Rubin LSST and to compute the flux observable through the photometric filter employed at the time of the detection. We are hence able to generate a simulation of an observation, as shown in Figure \ref{lc}. Out of all the simulated orphan afterglows, only approximately $\approx 4\%$ exhibited at least one point in their simulated observation.

\section{First version of a classifier}

\subsection{Characterisation of orphan afterglow light curves}

In order to describe as much as possible the shape of an ``observed" light curve, we first define several straightforward features: the duration between the first detection and the peak, the increase rate in the rising part of the light curve (if this rising part exists), the decrease rates in the first third and the last third of the decaying part, and the g-r magnitude colour. Only light curves with a minimum of 5 data points are considered for computing these features. To introduce additional features and help the classification, a fit of the light curve is performed using the function 
\begin{equation}
    ABMagnitude(t)=A \times t + B + C \times \exp(-D \times t)
\end{equation}
This function was determined with the assistance of \citep{russeil_2024}. To facilitate the fit, data points are rescaled to be on the r-band by using the fact that $F_\nu \propto \nu^{-\beta}$, with $\beta$ the spectral index. In our case, $\beta$ can take all the values between $-(p-1)/2$ when $\nu_m < \nu < \nu_c$ and $-p/2$ when $\nu_c < \nu$, $\nu_m$ and $\nu_c$ representing characteristic frequencies that decrease over time, and $p$ being the power-law index for the energy distribution of electrons. As the time of the burst for an orphan afterglow is not well constrained, the correct value of $\beta$ at the time of detection remains undetermined. We test the rescaling across several values of $\beta$ and select the one that minimises the distance between the rescaled points and the ``true" r-band points. To evaluate the fit quality, we also compute the $\chi^2$.

Each event is then described through the following features: the duration between the first detection and the peak, the increase rate, two decrease rates, the g-r colour, and parameters A, B, C, D, along with the $\chi^2$ value.

\subsection{Background data}

Orphan afterglows need to be discriminated from other transient events. To achieve this, we use the ELAsTiCC (Extended LSST Astronomical Time-series Classification Challenge) data set as source of many transients. This dataset, a DESC simulation of LSST alerts, aims to generate a realistic data stream for testing alert broker systems and classifiers. It contains synthetic transient light curves and host galaxies for different transient events (periodic and non-periodic) but does not include any GRB afterglows. For our classifier, we exclusively focused on non-periodic events, mainly supernovae, as periodic events are likely to be rapidly identified. Similar to orphan afterglows, the same set of features are computed for each of these events.

\begin{wrapfigure}[13]{r}{0.46\textwidth}
\vspace{-37pt}
    \begin{center}
        \includegraphics[width=0.46\textwidth]{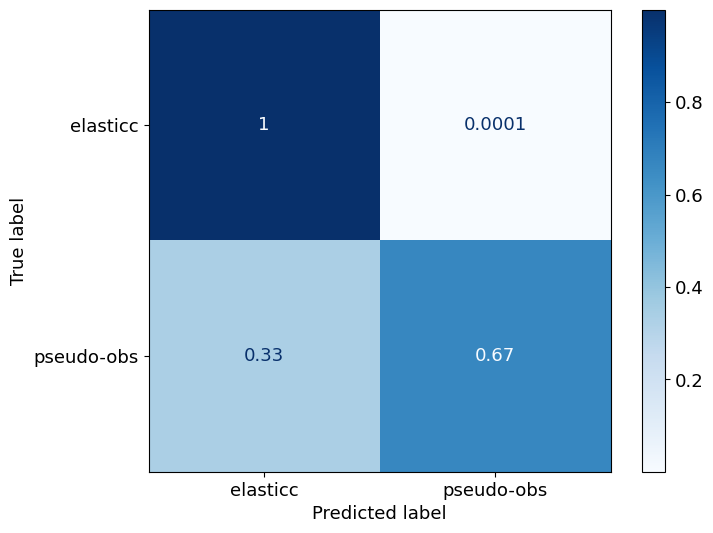}
        \captionsetup{width=0.43\textwidth}
        \caption{Confusion matrix of the classifier. The threshold to classify an event as an orphan is set to 0.999.}
        \label{matrix}
    \end{center}
\end{wrapfigure}

\subsection{Machine learning classifier}
 
We utilise the features for each class of event (orphan and ELAsTiCC) to test a Scikit-Learn gradient boosting classifier, a decision tree based machine learning algorithm. The classifier is trained on 500 orphans and 5000 ELAsTiCC events, and subsequently tested on 679 orphans and 10000 ELAsTiCC events. As shown on Figure \ref{matrix}, by retaining only events with a probability to be an orphan higher than 99.9\%, 452 out of 679 orphans and only 1 out of 10000 ELAsTiCC events remain. This is a promising start towards achieving our objective of obtaining a pure sample.

\section{Conclusion}

We are developing a machine learning classifier within the Fink framework to discriminate orphan GRBs afterglows among the Vera C. Rubin Observatory LSST data. 

To accomplish this, we simulated a population of short GRBs based on Swift BAT catalogues and their observations. We characterised the ``observed" light curves by defining specific features, which were used to train a machine learning algorithm to identify orphans, or at least exclude the majority of ELAsTiCC events. Preliminary testing and training resulted in the exclusion of nearly all ELAsTiCC events while retaining approximately two-thirds of the orphan afterglows, by setting a minimum probability threshold for an event to be classified as an orphan at 99.9\%. 

Improving this filter by incorporating additional statistics is a priority. Furthermore, we plan to adapt and test it using ZTF data, as Rubin LSST operations will start only in 2026. All the codes used in this work can be found in a GitLab repository\footnote{\href{https://gitlab.in2p3.fr/johan-bregeon/orphans}{https://gitlab.in2p3.fr/johan-bregeon/orphans}}.

%If you more commonly use the method of square brackets in the line of text
%for citation than the superscript method,
%please note that you need  to adjust the punctuation
%so that the citation command appears after the punctuation mark.

\scriptsize

\section*{Acknowledgments}

This work is carried on within the Rubin LSST Dark Energy Science Collaboration and is supported by the French National Research Agency in the framework of the "Investissements d’avenir” program (ANR-15-IDEX-02). The authors gratefully thank the Osservatorio Astronomico di Brera a Merate (INAF), Italy, for very useful discussions and comments on this project, especially M. G. Bernardini, P. D'Avanzo and O. S. Salafia. The authors also thank the IDEX for funding M. Masson's mobility grant. 

\setlength{\bibsep}{0pt plus 0.3ex}
\bibliography{ref}{}

\begin{thebibliography}{1}

\bibitem{Abbott_2017}
B.~P. Abbott et~al.
\newblock Multi-messenger observations of a binary neutron star merger.
\newblock {\em The Astrophysical Journal}, 848(2):L12, oct 2017.

\bibitem{Ghirlanda_2015}
G.~Ghirlanda et~al.
\newblock Unveiling the population of orphan gamma--ray bursts.
\newblock {\em Astronomy \& Astrophysics}, 578:A71, Jun 2015.

\bibitem{Moller_2020}
A.~Möller et~al.
\newblock Fink, a new generation of broker for the lsst community.
\newblock {\em Monthly Notices of the Royal Astronomical Society}, 501(3):3272–3288, November 2020.

\bibitem{Sari_1998}
R.~Sari et~al.
\newblock Spectra and light curves of gamma-ray burst afterglows.
\newblock {\em The Astrophysical Journal}, 497(1):L17–L20, Apr 1998.

\bibitem{Ryan_2020}
G.~Ryan et~al.
\newblock Gamma-ray burst afterglows in the multimessenger era: Numerical models and closure relations.
\newblock {\em The Astrophysical Journal}, 896(2):166, jun 2020.

\bibitem{D_Avanzo_2014}
P.~D’Avanzo et~al.
\newblock A complete sample of bright swift short gamma-ray bursts.
\newblock {\em Monthly Notices of the Royal Astronomical Society}, 442(3):2342–2356, June 2014.

\bibitem{Salafia_2023}
O.~S. Salafia et~al.
\newblock The short gamma-ray burst population in a quasi-universal jet scenario.
\newblock {\em Astronomy \& Astrophysics}, 680:A45, December 2023.

\bibitem{russeil_2024}
E.~Russeil et~al.
\newblock Multi-view symbolic regression, 2024.

\end{thebibliography}

\end{document}